\documentclass[letter,traditabstract]{aa}  
\usepackage[pdftex]{color,xcolor}
\usepackage[pdftex]{graphicx}
\usepackage{graphicx}
\usepackage{amstext} 
\usepackage{amsmath} 
\usepackage{amssymb} 
\usepackage[figuresright]{rotating}
\usepackage{natbib} 
\usepackage{afterpage}
\usepackage{txfonts}
\usepackage[urlcolor=blue,colorlinks=true,linkcolor=blue,citecolor=blue]{hyperref}
\usepackage{hyperref} 
\hypersetup{pdfauthor=Johannes Sahlmann}

\begin{document} 
\title{HD 5388 b is a $69\,\mathrm{M_{Jup}}$ companion instead of a planet}   
\author{J.~Sahlmann 
\and C.~Lovis
\and D.~Queloz
\and D.~S\'egransan}		
\institute{Observatoire de Gen\`eve, Universit\'e de Gen\`eve, 51 Chemin Des Maillettes, 1290 Sauverny, Switzerland\\
		\email{johannes.sahlmann@unige.ch}}			
\date{Received 18 January 2011 / Accepted 8 February 2011 } 

\abstract
{We examined six exoplanet host stars with non-standard Hipparcos astrometric solution, which may be indicative of unrecognised orbital motion. Using Hipparcos intermediate astrometric data, we detected the astrometric orbit of HD~5388 at a significance level of 99.4~\% ($2.7\,\sigma$). HD~5388 is a metal-deficient star and hosts a planet candidate with a minimum mass of $1.96\,M_\mathrm{J}$ discovered in 2010. We determined its orbit inclination to be $i =178.3^{+0.4\,\circ}_{- 0.7}$ and the corresponding mass of its companion \object{HD~5388~b} to be $M_2 = 69\pm20\,M_\mathrm{J}$. The orbit is seen almost face-on and the companion mass lies at the upper end of the brown-dwarf mass range. A mass lower than $13\,M_\mathrm{J}$ was excluded at the $3\,\sigma$-level. The astrometric motions of the five other stars had been investigated by other authors revealing two planetary companions, one stellar companion, and two statistically insignificant orbits. We conclude that HD~5388~b is not a planet but most likely a brown-dwarf companion. In addition, we find that the inclinations of the stellar rotation axis and the companion's orbital axis differ significantly.}
\maketitle
\section{Introduction}
For most extrasolar planets detected with the radial-velocity (RV) method, only their minimum masses are known because of their unknown orbit inclinations. Although the large number of planets allows us to draw statistically sound conclusions about their mass distribution \citep{Udry:2007sf, Howard:2010lr}, it is highly desirable to derive this distribution without the inclination incertitude. Astrometric measurements can resolve this ambiguity. For instance, \cite{Sahlmann:2011fk} used Hipparcos data to analyse the astrometric motion of stars with potential brown-dwarf companions and found that about half of the candidates are low-mass stars. Here, we extend this work to stars with planet candidates and non-standard Hipparcos solution, and discover the astrometric orbit of HD~5388, which hosts a recently announced planet candidate.
\begin{figure}\begin{center} 
\includegraphics[width= \linewidth, trim= 0 0.8cm 0 0cm,clip=true]{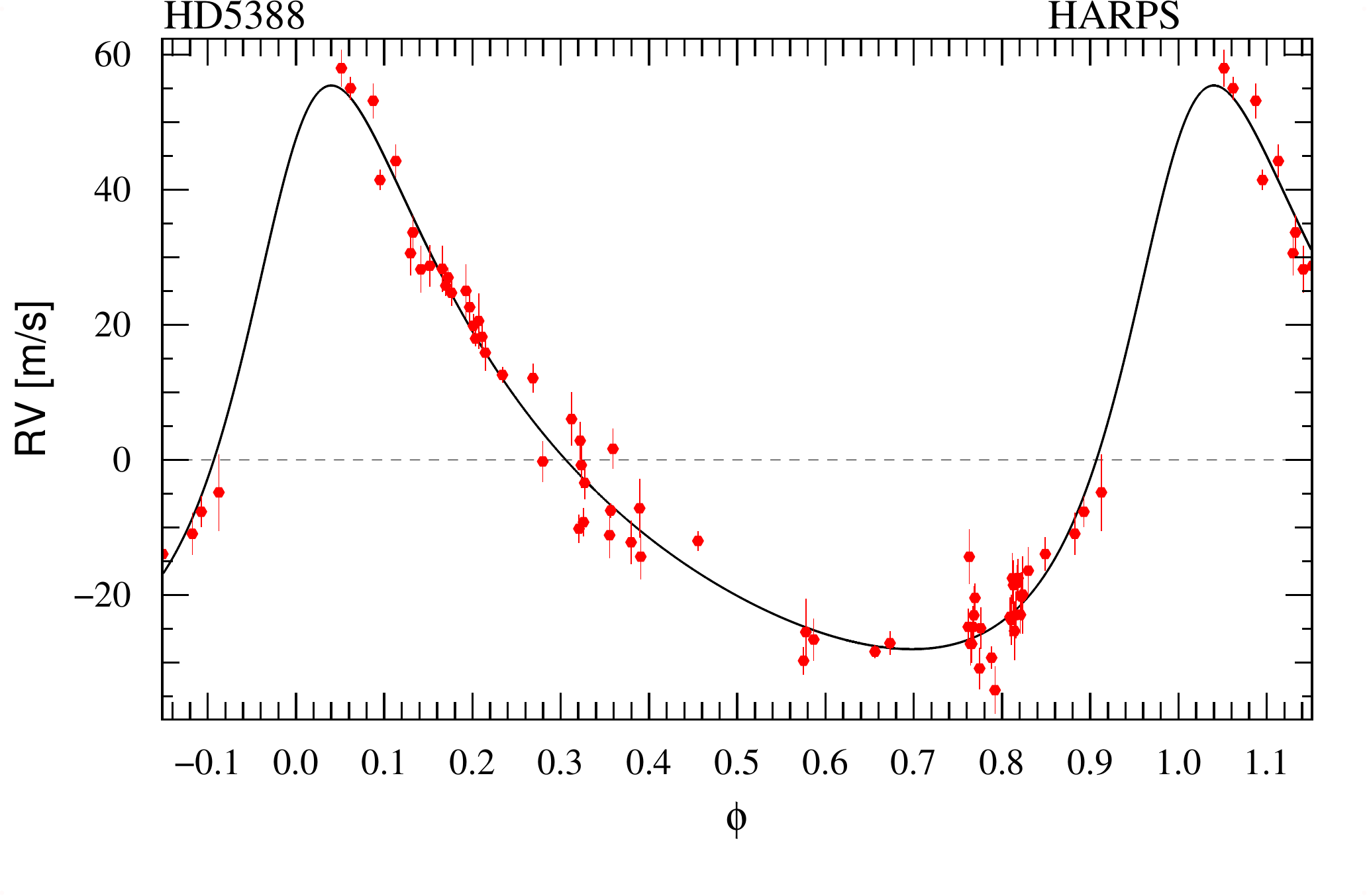} 
\caption{Phase-folded radial velocities of HD~5388 as published by \cite{Santos:2010fk2}. Red circles indicate the HARPS measurements and the solid line corresponds to the best-fit solution.}  \label{fig:rv}\end{center} \end{figure}
\section{Target selection and astrometric analysis}\label{sec:targets}
In September 2010, the list of RV-planets at \url{exoplanet.eu} contained 461 entries around 389 stars. Only 286 host stars are included in the new Hipparcos reduction \citep{:2007kx}. Many host stars of transiting exoplanets discovered in transit surveys (e.g., Kepler, CoRoT, HAT, WASP) and confirmed by RV are fainter than the completeness limit of Hipparcos. We selected the stars for which the new Hipparcos reduction found a non-standard solution and that were not flagged as member of a multiple system, thus having solution types '1', '7', or '9'. Type '1' solutions are termed stochastic and adopted when the standard five-parameter solution (type '5') is not satisfactory and neither orbital nor acceleration models improve the solution in terms of $\chi^2$. The types '7' and '9' are given, when the model has to include proper-motion derivatives of first and second order to obtain a reasonable fit. Six stars satisfied these criteria and are listed in Table~\ref{tab:4}.

The astrometric analysis was performed as described in \cite{Sahlmann:2011fk}, where a detailed description of the method can be found. We briefly recall the main elements of the analysis. Using the orbital parameters known from RV measurements, the intermediate astrometric data of the new Hipparcos reduction was fitted with a seven-parameter model depending on the inclination $i$, the longitude of the ascending node $\Omega$, the parallax $\varpi$, and offsets to the coordinates ($\Delta \alpha^{\star}$, $\Delta \delta$) and proper motions ($\Delta \mu_{\alpha^\star}$, $\Delta \mu_{\delta}$) given in the published catalogue of \cite{:2007kx}. A two-dimensional search grid in $i$ and $\Omega$ defined starting values for a standard nonlinear $\chi^2$-minimisation procedure identifying the global minimum. The uncertainties in the RV parameters are propagated to the astrometric solution by means of Monte Carlo simulations. The statistical significance of each astrometric orbit was determined by the distribution-free permutation test employing 1000 pseudo orbits. Uncertainties in the solution parameters were derived by Monte Carlo simulations. This approach has proven to be reliable in detecting orbital signatures in the Hipparcos astrometric data and efficiently distinguishing a significant orbit in the present low signal-to-noise ratio regime \citep{Sahlmann:2011fk}.
\begin{table}\caption{Orbital parameters of HD~5388 \citep{Santos:2010fk2}.}
\label{tab:1} 
\centering  
\begin{tabular}{l l c} 	
\hline\hline %
Parameter & Unit & Value \\
\hline
$P$ & (day) & $777.0 \pm   4.0$ \\
$e$ &  &$0.40 \pm 0.02$\\
$K_1$ & (ms$^{-1}$)  & $ 41.7 \pm   1.6$\\
$T_0$ &(MJD) &$54570.0 \pm    9.0$ \\
$\omega$  & (deg) &$324.0 \pm 4.0$ \\  
$M_2 \sin i$ &  ($M_\mathrm{J}$) &$ 1.96 $\\
$a \sin i$ & (mas)&0.05\\
\hline 
\end{tabular} 
\end{table}
\begin{table} 
\caption{Parameters of the astrometric solution for HD~5388.}
\label{tab:2} 
\centering  
\begin{tabular}{l l c} 	
\hline\hline %
Parameter & Unit & Value \\
\hline
$N_\mathrm{mes}$ & & 191\\
$N_\mathrm{orb}$ & &1.5\\
$\sigma_{\Lambda}$  & (mas)&3.8\\
$\chi^2_\mathrm{red}$ & &1.02\\
Significance & (\%) &99.4\\  
$\Delta \alpha^{\star}$ & (mas) &$-0.7 \pm0.6$\\
$\Delta \delta$ & (mas) & $0.9 \pm 0.6$  \\
$\varpi$ & (mas) & $18.9 \pm 0.7$ \\
$\Delta \mu_{\alpha^\star}$ & (mas $\mathrm{yr}^{-1}$)   & $0.5 \pm 0.6$\\
$\Delta \mu_{\delta}$ & (mas $\mathrm{yr}^{-1}$)   & $-0.1\pm0.5$  \vspace{1mm}\\
$\Omega$ &(deg) & $ 298.0^{+ 16.4}_{-26.5}$  \vspace{1mm}\\  
$i_\mathrm{orbit}$ & (deg)& $ 178.3^{+ 0.4}_{-0.7}$  \vspace{1mm}\\
$a$ & (mas)&$1.7\pm0.5$\\
$M_2$  &  ($M_\mathrm{J}$)& $ 69.2 \pm 19.9$\\ 
\hline 
\end{tabular} 
\end{table}

\section{The orbit of HD~5388}
\object{HD~5388} (HIP~4311) is listed with a stochastic solution in the catalogue of the new Hipparcos reduction. \cite{Santos:2010fk2} used the HARPS spectrograph to discover a massive planet candidate around this F6V-star. Figure~\ref{fig:rv} shows the radial velocities from the discovery paper. The companion has a minimum mass of $M_2 \sin i = 1.96\,M_\mathrm{J}$ and orbits its host in an eccentric orbit ($e=0.4$) with an orbital period of $P=777$ days. The time of periastron passage $T_0$, the longitude of periastron $\omega$, the radial-velocity semi-amplitude $K_1$, and the minimum semimajor axis of the star's astrometric motion $a \sin i$ are given in Table~\ref{tab:1}. The residuals of the RV solution ($3.3\,$ms$^{-1}$) are larger than the average measurement error ($2.8\, $ms$^{-1}$). By considering the star's spectral type and projected rotational velocity and analysing the bisector inverse slope, \cite{Santos:2010fk2} conclude that the excess RV noise is insignificant.

\begin{figure}\begin{center} 
\includegraphics[width= 0.9\linewidth, trim= 0 0cm 0 0cm,clip=true]{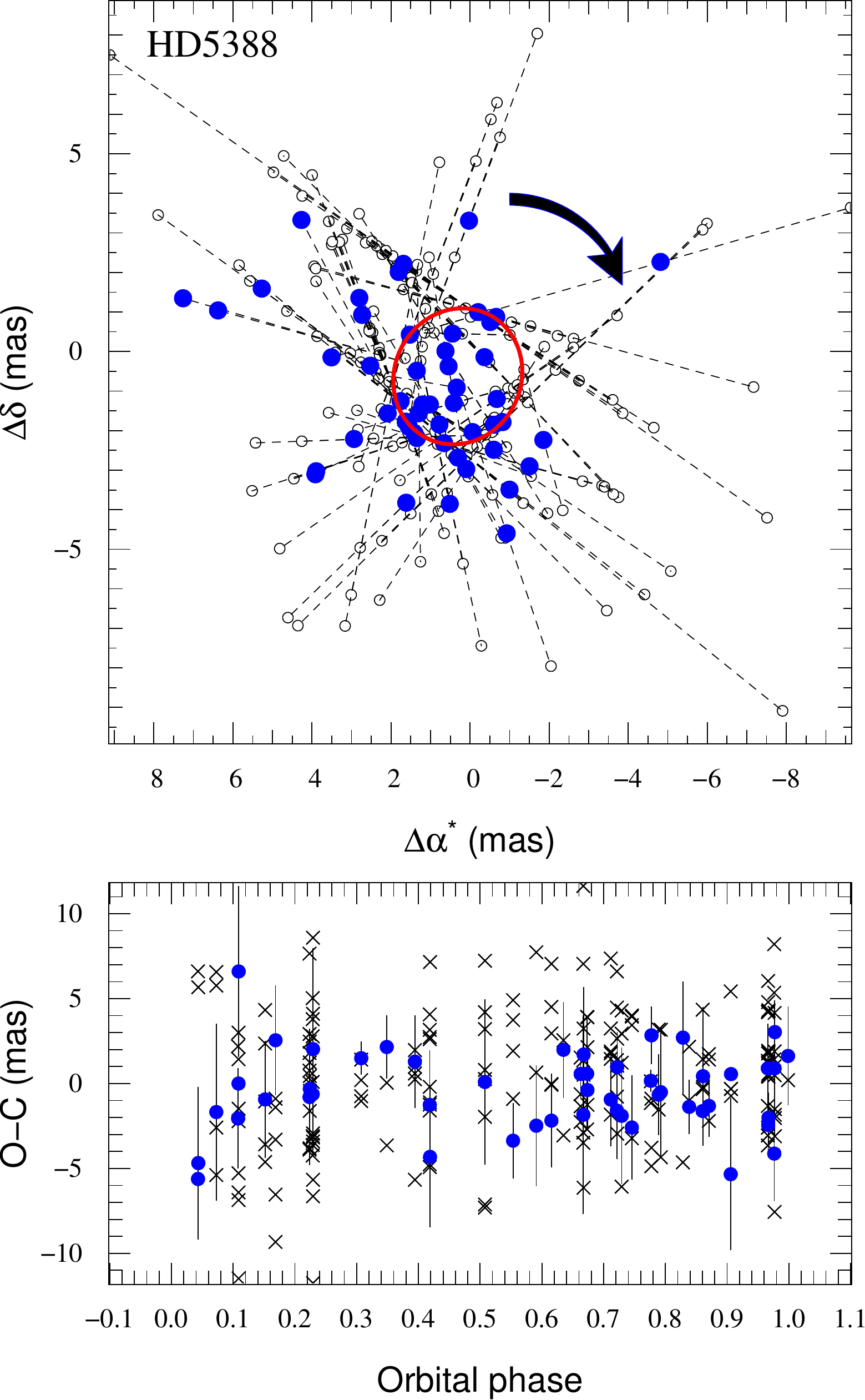}
 \caption{\emph{Top:} Astrometric orbit of HD~5388 projected on the sky. North is up and east is left. The solid red line shows the model orbit, which is orientated clockwise, and open circles mark the individual Hipparcos measurements. \emph{Bottom:} O-C residuals for the normal points of the orbital solution (filled blue circles) and the standard five-parameter model without companion (crosses).} 
\label{fig:orbits}
 \end{center} \end{figure}

We detected the astrometric orbit with a significance of 99.4~\% ($2.7\,\sigma$), which means that Hipparcos measured the stellar orbit induced by a companion with the orbital characteristics given in Table~\ref{tab:1}. The derived astrometric orbit with $a= 1.7\pm0.5$~milli-arcsec (mas) is small compared to the median single-measurement precision $\sigma_{\Lambda} = 3.8$~mas given by the new Hipparcos reduction. However, the large number of measurements $N_\mathrm{mes}=191$ covering more than one orbital revolution ($N_\mathrm{orb}=1.5$) resulted in an effective signal-to-noise ratio of $\mathrm{S/N}=a\cdot (\sigma_{\Lambda}/\sqrt{N_\mathrm{mes}})^{-1} = 6.2$ allowing the detection to be possible. With a reduced chi-square value of $\chi^2_\mathrm{red}=1.02$ the final seven-parameter fit is good and the derived offsets to positions, parallax, and proper motions are small. The values of $\Omega$ and the inclination $i_\mathrm{orbit}$ are well constrained at  $\Omega = 298^{+16\,\circ}_{-27}$ and $i_\mathrm{orbit} =178.3^{+ 0.4\,\circ}_{-0.7}$. Using a stellar mass of $1.2\,M_{\sun}$ \citep{Santos:2010fk2}, we derived the companion mass to be $M_2=69.2 \pm 19.9 \,M_\mathrm{J}$. The astrometric orbit of HD~5388 is shown in Fig.~\ref{fig:orbits} and its characteristics are summarised in Table~\ref{tab:2}. \cite{Reffert:2011fk} found an upper mass limit of $124\,M_\mathrm{J}$ for the companion of HD~5388, which is compatible with our result. 
 
Assuming that the orbit orientations are randomly oriented in space, the probability of measuring an inclination smaller than $2^\circ$ is only 0.06~\%. However, the number of planets detected in RV-surveys reached several hundreds making our detection reasonably probable. HD~5388 is a metal-deficient star with [Fe/H] = $-0.27\pm0.02$. The hydrogen-burning mass limit, commonly used to distinguish between stars and brown dwarfs, increases with decreasing metallicity and reaches $87\,M_\mathrm{J}$ ($0.083\,M_{\sun}$) for [Fe/H]~$<-1$ \citep{Chabrier:1997fr}. Because of the moderate metal-deficiency of HD~5388, we used a limit of $80\,M_\mathrm{J}$ and found that the probability that the companion of HD~5388 has a mass below this limit is 71~\%. The companion of HD~5388 is thus most likely a brown dwarf. This adds HD~5388 to the very few known Sun-like stars that have a brown-dwarf companion and a well-determined astrometric orbit.
\begin{figure}\begin{center} 
\includegraphics[width= \linewidth, trim= 0 0.5cm 0 0.5cm]{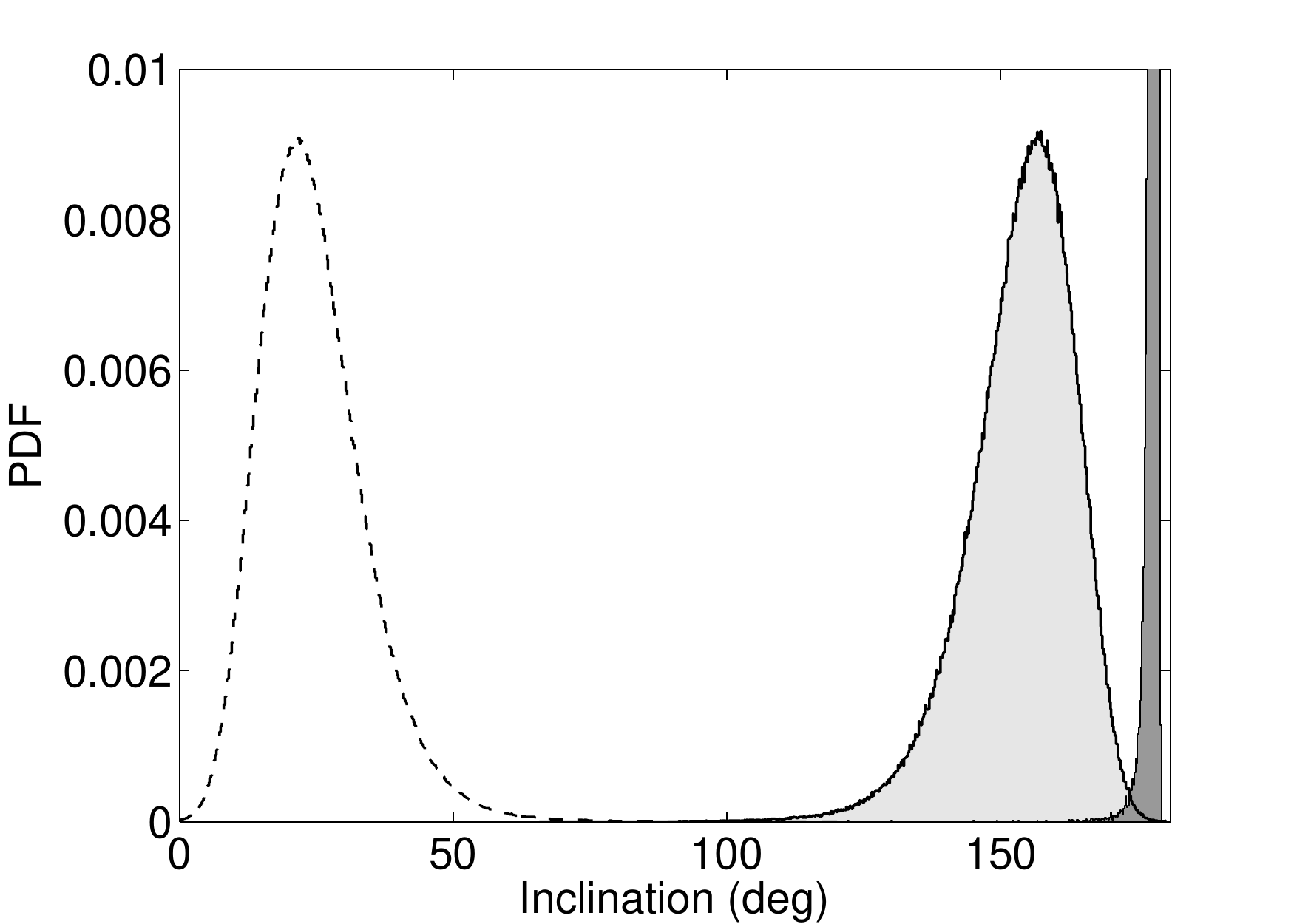}
 \caption{Probability density functions of orbit inclination ($i_\mathrm{orbit}$, dark-grey histogram) and stellar spin axis inclination ($i_\mathrm{rot}$, light-grey histogram). The orbit inclination PDF is very sharply peaked with its maximum at 0.18 and is truncated for display clarity. The PDF of $\psi = i_\mathrm{orbit}-i_\mathrm{rot}$ is shown as dashed line. The bin size is $0.2^\circ$.} 
\label{fig:inclinations}
 \end{center} \end{figure}

\subsection{Spin-orbit alignment}
After determining the orbital inclination from astrometry, we compared it to the orientation of the stellar spin axis. The inclination $i_\mathrm{rot}$ of the stellar spin axis is defined with respect to the line of sight ($i_\mathrm{rot}=0\degr$ or $180\degr$ for a pole-on view) and can be derived from the spectroscopic estimate of $ \upsilon \sin i_\mathrm{rot} = 4.2$~km~s$^{-1}$ \citep{Santos:2010fk2}. The authors did not give an error bar for this measurement and we assumed a conservative uncertainty of 1~km s$^{-1}$. On the basis of the activity indicator $\log R'_{H,K}$ determined from the HARPS spectra, we used the calibration of \cite{Mamajek:2008fk} to derive the stellar rotation period $P_\mathrm{rot}$. Assuming an effective temperature of $T_\mathrm{eff} = 6297 \pm 32$ \citep{Santos:2010fk2}, the apparent visual magnitude $m_V = 6.839 \pm 0.001$ \citep{:2007kx}, and the parallax $\varpi$ given in Table~\ref{tab:2}, we derived the star's absolute magnitude, luminosity ($L = 4.5 \pm  0.3\, L_{\sun}$), and radius ($R = 1.8 \pm  0.1 \,R_{\sun}$), using standard formulae and Monte Carlo resampling. Bolometric corrections were computed using the \cite{Flower:1996qy} parameters given by \cite{Torres:2010uq}. Using the stellar radius and rotation period, we derived the equatorial rotation velocity to be $\upsilon = 10 \pm 3$~km\,s$^{-1}$, and then the inclination of the spin axis by calculating $i_\mathrm{rot}  = \arcsin( \upsilon \sin i_\mathrm{rot} / \upsilon)$. 

Because we cannot determine the star's sense of rotation, the angle $i_\mathrm{rot}$ has a $180^\circ$ ambiguity and we first considered the value that falls into the same quadrant as $i_\mathrm{orbit}$, i.e. a prograde configuration. The astrometric analysis yielded the distribution of $i_\mathrm{orbit}$, obtained from 100\,000 Monte Carlo simulations. To obtain the $i_\mathrm{rot}$-distribution, we performed $10^6$ Monte Carlo simulations. We finally compared these two distributions by drawing $2\cdot10^7$ pairs of values $[i_\mathrm{orbit}$, $i_\mathrm{rot}]$ and obtained the distribution of the orbit obliquity or spin-orbit angle $\psi$ by computing $\psi = i_\mathrm{orbit} -i_\mathrm{rot}$. Figure~\ref{fig:inclinations} shows the probability density functions (PDF) of $i_\mathrm{orbit}$, $i_\mathrm{rot}$, and $\psi$.  The $i_\mathrm{orbit}$-distribution is very narrow and peaks around $ 178.3^\circ$, whereas the $i_\mathrm{rot}$-distribution is broad with its maximum at $ \sim\!155^\circ$ and the two distributions show a very small overlap. The median values and confidence intervals of $i_\mathrm{rot}$, $i_\mathrm{orbit}$, and the obliquity $\psi$ are given in Table~\ref{tab:spin}. We found that $\psi$ is larger than $4.4\degr$ and $19.3\degr$ with a probability of 99.7\% and 68.3~\%, respectively, and its nominal value with $1\,\sigma$ confidence intervals is $\psi= 23_{- 8}^{+10\,\circ}$ in prograde configuration, which indicates that the orientations of stellar spin and orbital axis differ substantially. In retrograde configuration, the obliquity would be $\psi'=153_{-10}^{+ 8\,\,\circ}$. 
\begin{table}\caption{Spin inclination, orbital inclination, and obliquity.}
\label{tab:spin} 
\centering  
\begin{tabular}{l l c l l} 	
\hline\hline %
Parameter & Unit & Value &  $1\,\sigma$ interval& $3\,\sigma$ interval\\
\hline
$\log R'_{H,K}$ &&$-5.00 \pm 0.02$& $\cdots$ & $\cdots$\\
$P_\mathrm{rot}$ & (day) & $ 9.4 \pm  2.0$ & $\cdots$ & $\cdots$\vspace{1mm}\\
$i_\mathrm{rot}$  & (deg) & $154.8_{-10.3}^{+ 8.1}$ & $(144.5, 162.9)$& $(111.1, 174.4)$ \vspace{1mm} \\ 
$i_\mathrm{orbit}$ & (deg) & $ 178.3_{- 0.7}^{+ 0.4}$ & $(177.6, 178.7)$& $(171.4, 179.1)$ \vspace{1mm} \\ 
$\psi$& (deg)  & $ 23.4^{+10.3}_{- 8.1}$ & $(15.2, 33.7)$& $( 3.3, 67.0)$ \vspace{1mm} \\
\hline 
\end{tabular} 
\end{table}
\begin{figure}\begin{center} 
\includegraphics[width= \linewidth, trim= 3.0cm 3.6cm 3.0cm 2.8cm,clip=true]{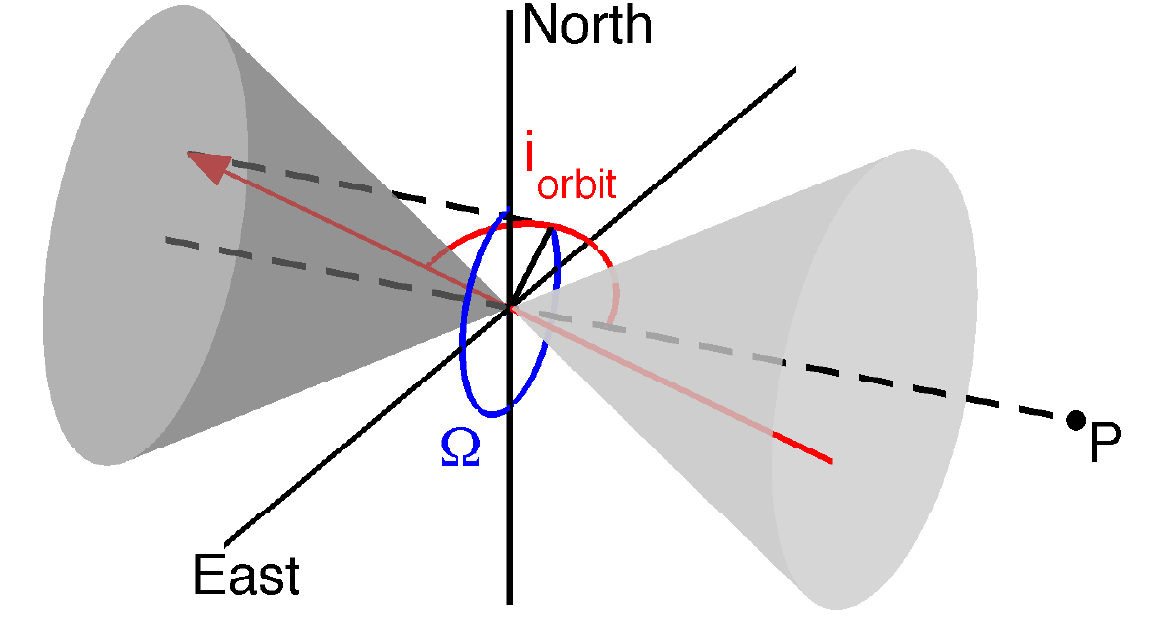}
\caption{Possible spin-orbit configurations. The astrometric motion in the sky plane defined by the north and east axes is observed from location P. The orbit orientation vector (red arrow) is defined by the angles $i_\mathrm{orbit}$ and $\Omega$. The stellar spin orientation vector can lie on either the dark-grey cone (prograde configurations) or the light-grey cone (retrograde configurations) with identical opening angles $2\cdot(180\degr -i_\mathrm{rot})=51\degr$.} 
\label{fig:inc}
 \end{center} \end{figure}

We note that these values of $\psi$ and $\psi'$ are lower limits, because the ascending node $\Omega_\mathrm{rot}$ of the spin axis is not constrained and we assumed that $\Omega = \Omega_\mathrm{rot}$ in their derivation. As illustrated in Fig.~\ref{fig:inc}, $\psi$ increases if the two ascending nodes $\Omega_\mathrm{rot}$ and $\Omega$ do not coincide, creating an additional uncertainty in $\psi$ of $2\cdot (180\degr-i_\mathrm{orbit}) = 3.4 \degr$ in the prograde configuration. For a retrograde orbit, this uncertainty is larger at $51 \degr$. In summary, the obliquity is $\psi= 23_{- 8}^{+10\,[+3.4]\,\circ}$ and $\psi'=153_{-10}^{+ 8\, [+51]\,\circ}$ for prograde and retrograde orbits, respectively, where the additional uncertainties are denoted in square brackets.

We performed these calculations using the $P_\mathrm{rot}$-calibration by \cite{Noyes:1984qy}, which yielded a value of $8.8 \pm  1.6$~days and an obliquity of $\psi= 22_{- 7}^{+9\,\circ}$, in agreement with the result obtained with the calibration of \cite{Mamajek:2008fk}. The isochrones of \cite{2008A&A...482..883M}\footnote{The interface is \url{http://stev.oapd.inaf.it/cgi-bin/cmd}.} indicated that HD~5388 is at the end of the core-H burning phase just past the turn-off point. Therefore, HD~5388 may be about to leave the main sequence or even be slightly evolved. This would introduce additional uncertainties and possibly biases into our estimates of $\log R'_{H,K}$, $P_\mathrm{rot}$, the stellar radius, and eventually $i_\mathrm{rot}$, and may therefore alter the above findings.

\section{The orbits of five additional selected stars}
In addition to HD~5388, we examined five other stars that satisfied our selection criteria in Sect.~\ref{sec:targets}. They are listed in Table~\ref{tab:4} and although their astrometric motions were previously studied in the literature, we describe our findings below:
\begin{itemize}
\item {55~Cnc} has five planetary companions announced by \cite{Fischer:2008it}. The planet 55~Cnc~d has a minimum mass of $M_2 \sin i = 3.84\,M_\mathrm{J}$. Using HST astrometry, \cite{McArthur:2004gd} derived an orbit inclination of $53\pm7^\circ$ for the d-planet, an astrometric perturbation size of $a = 1.9\pm0.4$~mas, and a mass of $M_2 = 4.9 \pm 1.1\,M_\mathrm{J}$. Compared to the measurement precision of $\sigma_{\Lambda} = 1.4$~mas, the stellar orbit is large enough to have been measured by Hipparcos, which explains the non-standard solution type. Because the Hipparcos data only covers $20$~\% of the 5200~day orbital period, we did not attempt to fit an astrometric orbit. 
\item{HD~81040} hosts a massive planet with $M_2 \sin i = 6.86\,M_\mathrm{J}$ \citep{Sozzetti:2006fj}. We did not detect orbital signature because the obtained orbit has a very low significance of $68~\%$, in agreement with the result of \cite{Sozzetti:2006fj}. As in that work, we used the Hipparcos astrometric data to set an upper limit to the companion mass of $47.5\,M_\mathrm{J}$, thus excluding that the companion has a stellar nature.
\item{HD~179949} hosts a planet candidate with $M_2 \sin i = 0.84\,M_\mathrm{J}$ \citep{Tinney:2001yq}. We did not detect an orbital signature because the obtained orbit has a very low significance of $71~\%$, which confirms the result of \cite{Zucker:2001ve}.
\item{HD~195019}  hosts a planet candidate with $M_2 \sin i = 3.70\,M_\mathrm{J}$ \citep{Wright:2007qy}. We detected the astrometric orbit with a high significance of $98~\%$ and a very low inclination of $0.3\pm0.1^\circ$, which renders a stellar companion and confirms the result of \cite{Zucker:2001ve}.
\item{$\gamma$ Cep} (\object{HD~222404}) hosts a planet candidate with $M_2 \sin i = 1.7\,M_\mathrm{J}$ \citep{Hatzes:2003lr}. We did not treat this system because an exhaustive analysis, including the use of Hipparcos astrometry, was performed by \cite{Torres:2007lr}, who set an upper mass limit of $16.9\,M_\mathrm{J}$ to the planetary companion.
\end{itemize}
\begin{table}\caption{Stars with candidate RV-planets and non-standard Hipparcos astrometric solutions.}
\label{tab:4} 
\centering  
\begin{tabular}{r r r c r r r r} 	
\hline\hline %
Object & HD & HIP    & Sol.& Sp. T. & \\  
       	&       & 	    &type &                            \\  
\hline 
HD~5388 & 5388 &4311 & 1 & F6V  \\ 
\object{55~Cnc} & 75732 &  43587 & 1 & G8V \\			
\object{HD~81040} &81040& 46076& 7 & G0V \\
\object{HD~179949} & 179949 & 94645 & 1&   F8V   \\ 
\object{HD~195019} & 195019 &100970 &  7&  G3IV \\ 
$\gamma$~Cep & 222404 &  116727 & 1 & K1IV \\			
\hline
\end{tabular} 
\end{table}
\section{Discussion}
Among the potential planets around the six stars selected in Sect.~\ref{sec:targets}, one turned out to be a star (HD~195019~b), two are likely planets (55~Cnc~d and $\gamma$~Cep~b), and one is a brown-dwarf companion (HD~5388). The astrometric orbits of two stars were not detected with Hipparcos and the companions remain planetary candidates. The significant fraction of one third of non-planetary companions found in this small sample is a selection effect and does not reflect the properties of the planetary population. That a large population of non-planetary companions is observed with low orbital inclinations, hence mistaken for planets, is statistically unlikely and would have been detected with Hipparcos \citep{Zucker:2001ve}. At the individual level, however, a planet detected by RV remains a planet \emph{candidate} until its mass range is confined by complementary constraints derived for instance from a transit lightcurve, dynamical interactions, by direct imaging, or from astrometric data, as in the case of HD~5388. In the future, as the data of the GAIA astrometry satellite becomes available, we will be able to measure the astrometric orbits of many planetary systems, thus obtaining an accurate census of the planetary mass distribution. A few more planet candidates may then turn out to be brown-dwarf companions.           

We have detected the astrometric orbit of HD~5388, which harboured a planet candidate of minimum mass $M_2 \sin i = 1.96\,M_\mathrm{J}$. We found  an almost face-on orbit with an inclination of $i_\mathrm{orbit}= 178.3_{- 0.7}^{+ 0.4\,\circ}$. Consequently, we determined the mass of the companion HD~5388~b to be $M_2 = 69\pm20\,M_\mathrm{J}$. A mass lower than $13\,M_\mathrm{J}$ was excluded at the $3\,\sigma$ confidence level. Thus, HD~5388~b can no longer be considered a planet. It is instead a close brown-dwarf companion, which is remarkable because at most $0.6~\%$ of Sun-like stars have close brown-dwarf companions \citep{Sahlmann:2011fk}. In particular, it is one of the very few brown-dwarf companions for which the astrometric orbit could be measured and whose orbital motion is therefore fully characterised, such as \object{HD~38529} \citep{Benedict:2010ph}. 
 
The measurement of the spin-orbit alignment in planetary and binary systems, e.g., \cite{Triaud:2010qy} and \cite{Albrecht:2011uq}, gives insight into the formation of these systems. The most frequently applied technique uses spectroscopic measurements during the transit of eclipsing systems and only determines the projected spin-orbit angle. Because it is restricted to eclipsing systems, it favours small-separation, short-period systems (another method relies on interferometric imaging of the stellar surface, see \citealt{2011ApJ...726..104H}). We determined the stellar spin axis inclination on the basis of the spectroscopic estimate of $\upsilon \sin i_\mathrm{rot}$ and the rotation period calibration using $\log R'_{H,K}$. We found evidence of non-zero spin-orbit angles with numerical values of $\psi= 23_{- 8}^{+10\,[+3.4]\,\circ}$ and $\psi'=153_{-10}^{+ 8\, [+51]\,\circ}$ for prograde and retrograde orbits, respectively. Thus, for both orientations of the unknown stellar spin, we found substantial non-coplanarity between the stellar and orbital axes. Spin-orbit angle measurements obtained from the determination of the astrometric orbit and the spectroscopic $\upsilon \sin i$ offer the opportunity to access this information for non-eclipsing systems.
\begin{acknowledgements}
J. S. thanks A. Triaud for his contributions to the discussion on spin-orbit angles. 
\end{acknowledgements}
\bibliographystyle{aa} 
\bibliography{16533} 
\end{document}